\documentclass{article}

\usepackage{arxiv}

\usepackage[utf8]{inputenc} 
\usepackage[T1]{fontenc}    
\usepackage{hyperref}       
\usepackage{url}            
\usepackage{booktabs}       
\usepackage{amsfonts}       
\usepackage{nicefrac}       
\usepackage{microtype}      
\usepackage{lipsum}
\usepackage{cite}
\usepackage{amsmath,amssymb,amsfonts}
\usepackage{algorithmic}
\usepackage{graphicx}
\usepackage{textcomp}
\usepackage{epstopdf}
\usepackage{xcolor}
\usepackage{subfig}
\usepackage{siunitx}
\usepackage{textcomp}
\usepackage{amsmath}

\title{Tunable THz and Infrared Plasmonic Filters and Switches Based on Circular Graphene Resonator with 90\pmb{$^\circ$} Bending of Output Port}

\author{
  Victor Dmitriev\thanks{Use footnote for providing further
    information about author (webpage, alternative
    address)---\emph{not} for acknowledging funding agencies.} \\
 Department of Electrical Engineering\\
  Federal University of Para\\
  66075-900, Belem, Para, Brazil. \\
  \texttt{victor@ufpa.br} \\
   \And
Geraldo Melo\\
  Federal Rural University of Amazonia\\
 66.077-830, Belem, Para, Brazil.\\
    \texttt{geraldo.melo@ufra.edu.br} \\
  \AND
  Wagner Castro\\
  Institute Cyberspace\\
  Federal Rural University of Amazonia\\
 66.077-830, Belem, Para, Brazil.\\
    \texttt{wagner.ormanes@ufra.edu.br} \\
}

\begin{document}
\maketitle

\begin{abstract}
Two types of novel graphene-based components, namely,  filters and  electro-optical switches in guided wave configuration are suggested and analysed. The filters  differ from the known ones with collinear orientation of the input and output waveguides by geometry and symmetry. The components consist of a  circular graphene disk and two  nanoribbons oriented at $90^\circ$ to each other in the plane of the graphene layer. The graphene elements  are placed on a dielectric substrate. We show that change in symmetry leads to a drastic change in the properties of the components. The physical principle of the devices is based on the dipole, quadrupole and hexapole resonances in the graphene disk which define stop-band, pass-band and stop-band frequency characteristics, respectively. A combination of stop-band and pass-band filter properties by shifing the electronic Fermi level allows one to design a switch. Numerical simulations show that the suggested components have  very  small dimensions, good characteristics and provide a dynamic control of their central frequency via electrostatic gating.

\end{abstract}


\section{Introduction}
Graphene, which is the first two-dimensional material \cite{Geim}, possesses very special electromagnetic properties and can be used for generation, detection and manipulation of THz and infra-red electromagnetic spectrum, in particular in sensors, switches, modulators, filters, among  others. Surface plasmon-polaritons (SPPs) propagate in graphene with better characteristics than in metals. Parameters of graphene can be easily adjusted dynamically.

Switch is one of the key elements of  digital technology. The physical principles used in switches depend on the frequency region, on the power of electromagnetic waves and on the required time of switching. In optics, for example,  electro-optical, thermo-optical, acousto-optical and non-linear effects are used. In particular, electro-optical Mach-Zehnder interferometer is a typical elements in optical circuits \cite{Mach-Zehnder}. In photonic crystal technology, many types of switches were also suggested \cite{photonic-crystal}. 

 An important advantage of the graphene switches is their sub-wavelength dimensions because the wavelength of SPP $\lambda_{SPP}$ is one-two orders smaller that the free space wavelength $\lambda_{0}$. In \cite{Liu}, graphene-based dielectric waveguide optical modulator is presented. A combination of photonic crystal cavity with graphene to provide electrical control is discussed in \cite{Maj}. In \cite{Kim}, a hybrid graphene-gold nanorod system for near-infrared region is suggested which can be used as a modulator. The electro-optical switches have  advantages in comparison with other types of switches in terms of low power consumption and high switching speed. In \cite{z}, an infra-red switch based on graphene waveguide strip is considered.  The switching mechanism is similar to the electro-optical components in the optical region. In order to obtain a sufficient isolation level, the section of the switch with OFF state electrostatic field in the distributed switch should have the length of several wavelength of SPP waves.
 
 In the integrated networks, one of the principal aims is to achieve a high density of elements. A significant reduction of the switch dimensions to  $(0.5\div1) \lambda_{SPP}$ can be achieved by using the structures based on resonance effects. The resonant cavities increase the time of wave interaction with graphene enhancing the necessary effects. The resonances in circular graphene cavity with the lateral coupling have been studied theoretically in \cite{Zhang} and with the front coupling in \cite{filter}. The existence of the dipole, quadrupole, hexapole and other modes in circular graphene resonators have been confirmed experimentally in \cite{Nikitin}.  

 Below, we discuss two new components for THz and  far-infrared  regions, namely, a filter with stop-band or pass-band properties and a new ultra-compact electro-optical  switch. The analysis will be fulfilled by using circuit theory, group theory and full-wave calculus. We will show that these devices based on a circular graphene resonator, have a very simple structure, small dimensions  and good characteristics.
\section{Description of Design, Symmetry and Materials}
\begin{figure}[t]
\centering
\includegraphics[width=28pc]{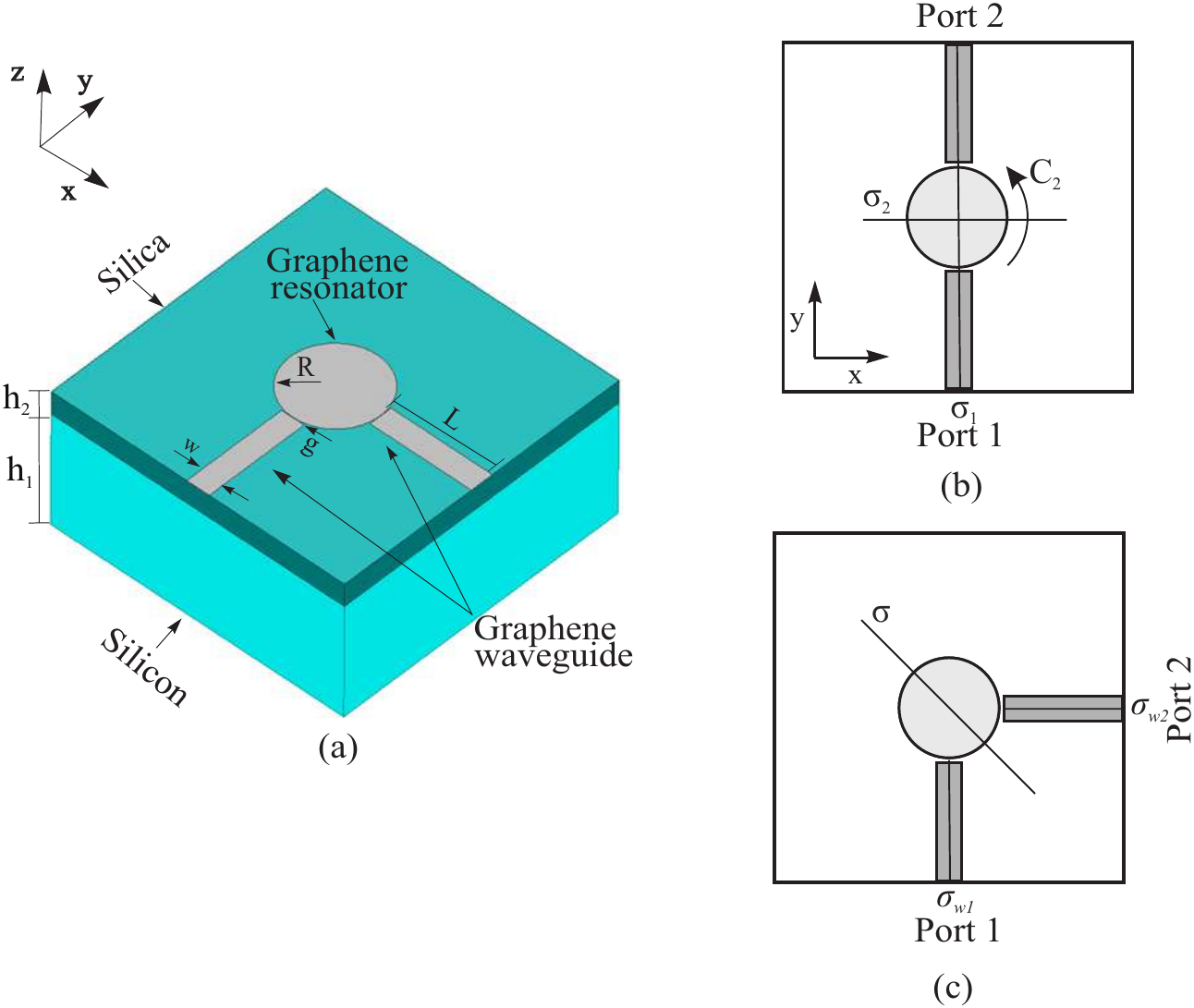}
\caption{(a) Schematic representation of suggested graphene filter and switch; (b) Front view of devices collinear  filter \cite{filter} with the symmetry planes $\sigma_1$, $\sigma_2$ and the two-fold axis of rotation $C_2$; (c) filter (switch) with 90$^{\circ}$ bending described by the plane $\sigma$ (global symmetry) and two planes $\sigma_{w1}$ and $\sigma_{w2}$ of the waveguides (local symmetry). }
\label{fig_Fig12}
\end{figure}

 The proposed device is demonstrated schematically in Fig.\ref{fig_Fig12}a,c. For comparison, the known filter \cite{filter} is depicted in Fig.\ref{fig_Fig12}b. Both devices consist of a circular graphene resonator with the radius $R$ and two graphene strips with the width $w_i$ (i = 1, 2) and the length $L$ coupled frontally to the resonator with a small gap $g$. The two strip waveguides in the known filter are collinear, but in the suggested structure they are oriented at 90$^{\circ}$ to each other. The graphene elements placed on a dielectric substrate, are situated in the plane x0y. The dielectric layers of silica SiO$_{2}$ and silicon  Si have the  thickness $h_{2}$ and $h_{1}$  and the relative permittivity $\varepsilon_1$ and  $\varepsilon_2$, respectively.

 The collinear filter in Fig.\ref{fig_Fig12}b described by the group $C_{2v}$ (we use the Schoenflies notations of the groups \cite{victor}) possesses two symmetry planes $\sigma_1$, $\sigma_2$ and the two-fold rotation around the axis z denoted by $C_2$. In the new structure with the 90$^{\circ}$ orientation of the input and output strips, only the plane of symmetry $\sigma$ exists, i.e. the group is $C_s$. The irreducible representations IRREPs of the group $C_{s}$ \cite{victor} are given in Table  of Appendix, where IRREP $A$ corresponds to the even mode and $B$ to the odd mode with respect to the plane  of symmetry $\sigma$ in the resonator (see the third column of this Table). 
 
 In contrast to the collinear structure, the planes of symmetry of the nano-strips $\sigma_{w1}$ and $\sigma_{w2}$ (local symmetry) in the new component do not coincide with the plane of symmetry of the whole structure $\sigma$ (global symmetry), see Fig.\ref{fig_Fig12}c. The full symmetry of the structure in this case is defined by the direct product $C_s \otimes C_s$. 
 Including the local symmetry in the description allows one to obtain an additional information about the properties of the structure. A discussion of the corresponding mathematics  is out of scope of this paper. Notice only, that this problem is related to the theory of point defects in crystals which is presented  in \cite{Evar}. We give a schematic illustration of the global $C_s$ and local $C_s \otimes C_s$ symmetries in our structure in Fig.\ref{fig_Appendix2} of Appendix. One can see in this figure a symmetry of the electric field $|E_z|$ in the region of junction of the waveguides with the resonator which can appear due to local $\sigma_{wi} (i=1, 2)$ symmetry of the waveguides. In Fig.\ref{fig_Appendix2}c, the whole structure of the field is even with respect to $\sigma$, but the local field around the waveguide-resonator junction is odd with respect to $\sigma_{wi}$. In Fig.\ref{fig_Appendix2}d, the whole structure is odd with respect to $\sigma$, and the local field is also odd with respect to $\sigma_{wi}$.

 To calculate the electrical conductivity of the graphene, we use semi-classical model of Drude described in \cite{inter}:
\begin{equation}
\sigma = \frac{2D}{\pi} \frac{i\omega - 1/\tau}{( \omega + i / \tau ) ^2},
\label{eq1}
\end{equation}%
where $D = 2\sigma_0 \epsilon _F / \hbar$ is the Drude weight, $\sigma_0$ is the universal conductivity, $\epsilon_F$ is Fermi energy of graphene, $\hbar$ is the reduced Planck's constant, $e$ is the electron charge, $\tau$ is the relaxation time and $\omega$ is the frequency of the incident wave. In the following, we shall use the value of $\tau=0.5$ ps, which is reasonable for the case of graphene on SiO$_2$ substrate. Notice also, that by electrostatic doping  graphene Fermi level $\epsilon_F$ of 1.2 eV  can be reached \cite{Khrap}. 

The numerical simulations were performed using the full-wave  commercial software  HFSS \cite{HFSS} which is based on the method of finite elements. Here,  graphene was modelled as an infinitesimally thin sheet using the surface impedance boundary conditions \cite{Transformation}.  
\section{Graphene Plasmonic Strip Waveguide and Resonator}
 It is well known that graphene strips can support two kinds of guided SPP modes, namely  the fundamental mode and edge modes which are discussed in details in  \cite {HE2013}, \cite{Edge}. We choose the fundamental symmetric mode, which has more uniform distribution of the fields along the width of the strip in comparison with the higher modes. Symmetry of the electric field $|E_z|$ of this mode is even with respect to the plane of symmetry of the strip $\sigma_{wi}$, i.e. this field component belongs to the IRREP $A$ of Table  in Appendix. 
 
 In order to exclude the influence of the lengths of the  strip waveguides on the losses of the switch, we will consider in the frequency characteristics of the filters and switches  only the losses related to the disk resonator. For this purpose, we subtract from the total losses of the device the losses of the waveguide with the length $2L$, see Fig.\ref{fig_Fig12}a. 
 
 The infinite graphene placed on a dielectric substrate supports transverse-magnetic (TM) SPP waves with the dispersion relation given by \cite{Livro}:
\begin{equation}
\beta_{spp}=\dfrac{(1+\varepsilon_1)(\omega\hbar)^2}{4\alpha\epsilon_F\hbar c},
\label{eq2}
\end{equation}%
where $\beta_{spp}$ is SPP propagation constant, $\alpha=e^2/(4\pi \varepsilon_0 \hbar c) \approx 0.007$ is the fine-structure constant and $\varepsilon_1$ is the dielectric constant of the substrate. 

\begin{figure}[t]
\centering
\includegraphics[width=22pc]{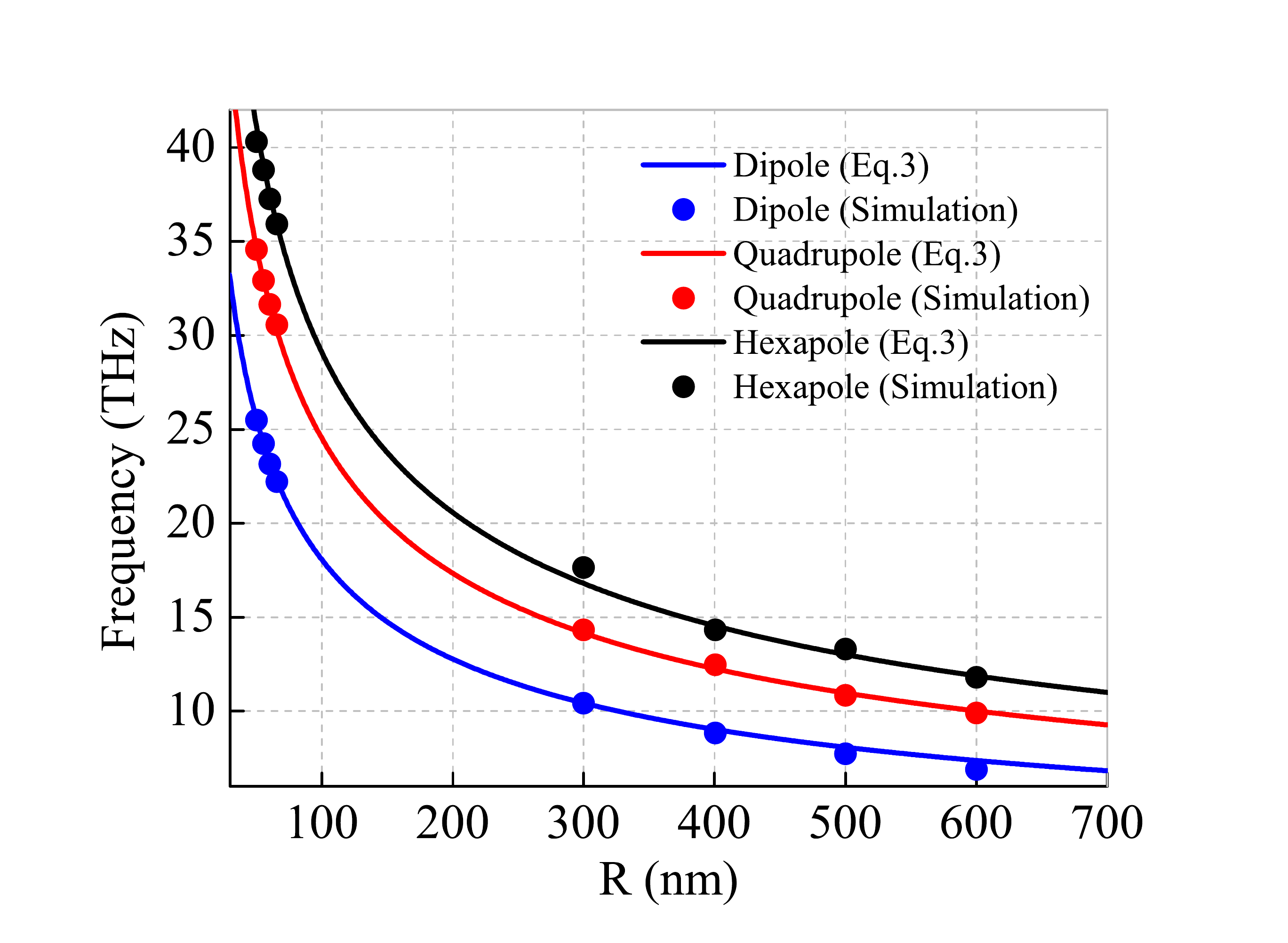}
\caption{Resonant frequency of resonator versus radius $R$, $\epsilon_{F} = 0.3$ eV.}
\label{fig_RaioEq}
\end{figure}

It is known \cite{borda} that the structure of the fields in the resonator corresponds to edge-guided waves. Thus, one can define the radius $R$ of the resonator with dipole, quadrupole or hexapole  mode from the condition of edge-guided mode resonance $2\pi R = n\lambda_{spp}$, where $\lambda_{spp}$ is the wavelength of the SPP mode and n = 1, 2, 3 is the mode number. From the relations $\beta_{spp}=2\pi/\lambda_{spp}$ and $R=n\lambda_{spp}/2\pi$, one obtains:
\begin{equation}
R\approx na_nA\dfrac{\epsilon_F}{(1+\varepsilon_1)\omega_{c}^2},
\label{eq3}
\end{equation}%
where $A = 8.3\times10^{40} (kg.m)^{-1}$, $\epsilon_{F}$ is Fermi energy (given in $J$) and $a_n$ is the correction coefficient for $n$ mode, $a_1=1, a_2=0.96, a_3=0.93$ for the dipole, quadrupole or hexapole mode, respectively. One can see that the radius of the resonator $R$ (given in $m$) depends on the central frequency of the filter and switch $\omega_{c}$ (in rad/s), on $\epsilon_F$ and $\varepsilon_1$. The central frequencies of the devices for the modes are increased with decreasing the radius $R$ in accordance with  the relation $\omega_{c}\propto1/\sqrt{R}$ as it follows from (\ref{eq3}). Fig.\ref{fig_RaioEq} demonstrates a very good agreement between the analytical and numerical results.
\section{Resonant Frequency versus Fermi Energy}
\begin{figure}[t]
	\centering
	\includegraphics[width=22pc]{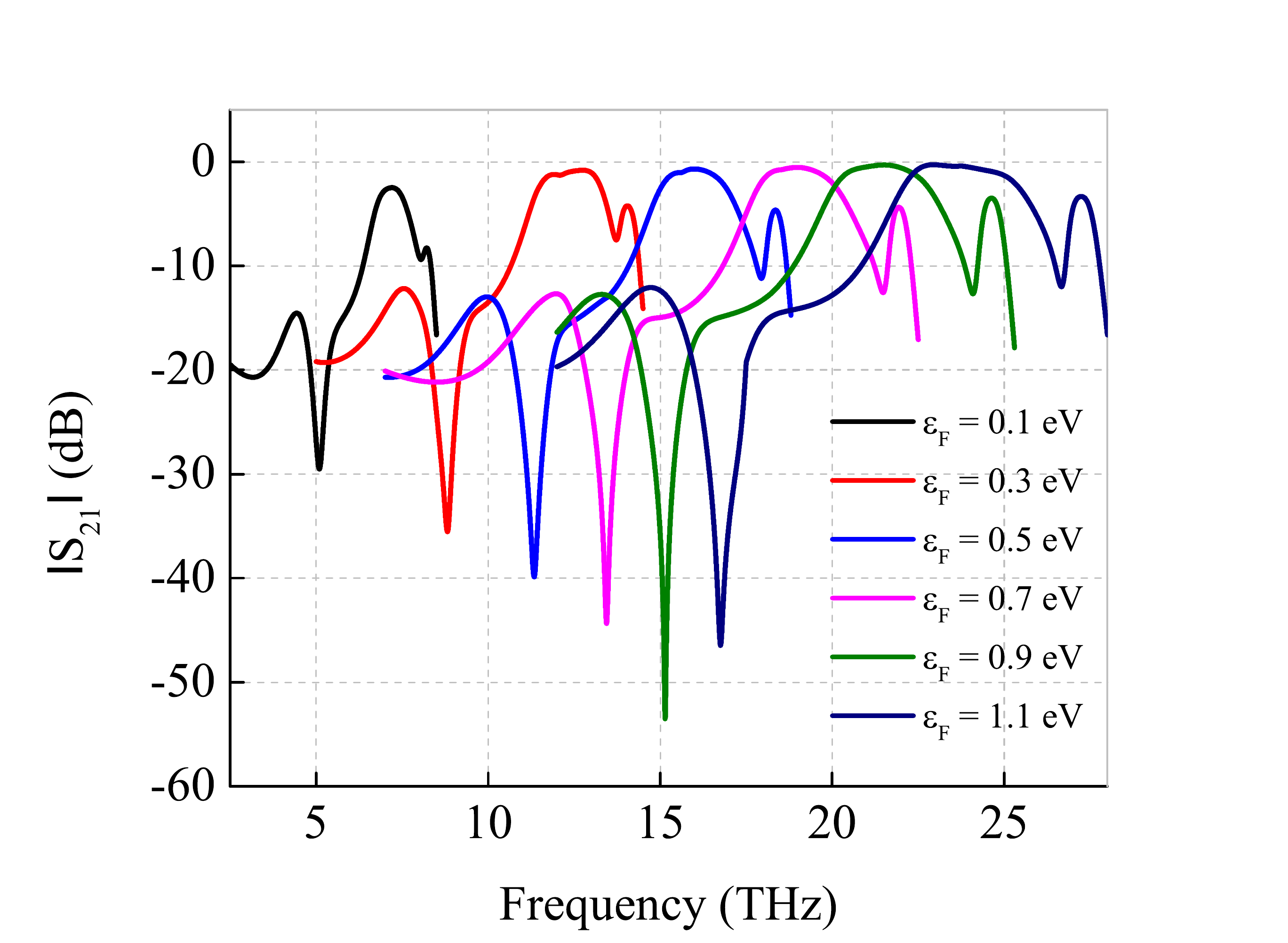}
	\caption{Frequency responses of resonator for different values of $\epsilon_F$, $g$ = 2 nm, $w$ = 200 nm, $L$ = 900 nm, $R$ = 400 nm and $\tau$ = 0.5 ps.}
	\label{fig_Mi2}
\end{figure}

 The influence of the Fermi energy on the resonance frequencies of  the structure  was investigated. The Fermi energy was varied from 0.1 eV to 1.1 eV and the corresponding frequency responses are plotted in Fig.\ref{fig_Mi2}. It can be observed that the increase of the Fermi energy shifts the resonant frequency to higher values. With change of $\epsilon_F$ between 0.1 eV and 1.1 eV, the resonant frequency ranges from 5 THz to 15.96 THz for the dipole mode, from 7.2 THz to 22.56 THz for the quadrupole mode and from 8.2 THz to 25.9 THz for the hexapole one.
\section{Circuit Theory of Switch}
 In microwave circuit theory \cite{Altman}, \cite{POZAR} a linear, time-invariant and passive two-port is described in general by a $2\times2$ scattering matrix [S]. Below, we shall discuss S-parameters of the filters and switches, which describe the transmission and reflection properties of the devices:
\begin{equation}
[S]=\left(\begin{array}{ccc}
	 S_{11} & S_{12}\\ 
	 S_{21} & S_{11}\end{array} \right),
\label{eq4}
\end{equation}%
where the matrix $[S]$ is defined in terms of incident and reflected voltage waves. 

In this Section, we shall use the unitary constraint  on the scattering matrix $[S]([S]^{\ast})^t = [I]$, where $[I]$ is the $2\times2$ unit matrix, $^\ast$ means complex conjugation and $^t$ denotes transposition. This constraint means that the device is lossless. Reciprocity of the devices imposes the restriction on the matrix elements $S_{21} = S_{12}$ \cite{victor} where $S_{21}$ is the transmission coefficient from port 1 to port 2 and $S_{12}$ is the transmission coefficient from port 2 to 1. Another restriction comes from the  the plane of symmetry $\sigma$, see Fig.\ref{fig_Fig12}c. Due to this plane of symmetry  one has $S_{22} = S_{11}$, where $S_{11}$ and $S_{22}$ are the reflection coefficients at port 1 and 2, respectively.   Now we shall consider separately the two states of the switch: state $\it OFF$ and state $\it ON$.
\subsection{State $\it OFF$ of Switch}
 The scattering matrix for the ideal switch in the state $\it OFF$ is
\begin{equation}
[S]_{OFF}=\left(\!
\begin{array} {cc}
1   & 0    \\
0   & 1
\end{array} \!\!
\right),
\label{eq5}
\end{equation}%
i.e. we suppose that the  physical principle in this state is based mostly on the  reflection of the incident wave. 
The matrix $[S]$ eigenvalue problem is defined by equation $[S]{\bf V}=s{\bf V}$ where $\bf V$ is an eigenvector and $s$ is the corresponding eigenvalue. The matrix $[S]_{OFF}$ is  the unit one which has two equal eigenvalues $s_{1OFF}=s_{2OFF}=1$. A peculiarity of $[S]_{OFF}$ is that  any vector can serve as an eigenvector ${\bf V}_{1OFF}$ and ${\bf V}_{2OFF}$. For example, one can choose
\begin{equation}
{\bf V}_{1OFF} =\frac{1}{\sqrt{2}}\left(\!
\begin{array}{c}
1 \\
1
\end{array}\!
\right), \qquad {\bf V}_{2OFF} = \frac{1}{\sqrt{2}}\left(\!
\begin{array}{c}
1 \\
\!\!\!-1
\end{array}\!
\right),
\label{eq6}
\end{equation}%
where $1/\sqrt{2}$ is the normalization coefficient. 

Independently on the structure of the matrix $[S]$, the symmetry group $C_{s}$ indicates two types of possible excitations of the two-port: the even excitation corresponding to the IRREP $A$ and the odd one corresponding to IRREP $B$ of Table in Appendix. The even and odd excitations are described by the vectors ${\bf V}_e$ and  ${\bf V}_o$ with the  eigenvalues $s_{e}=s_{o}=1$. The excitation of only port 1 can be obtained by the sum of the even and odd  excitations (${\bf V}_e + {\bf V}_o$) which is also an eigenvector of matrix $[S]_{OFF}$ with the same eigenvalue $s=1$.
\subsection{State $\it ON$ of Switch}
 In the state $\it ON$, the ideal scattering matrix is
\begin{equation}
[S]_{ON}=\left(\!
\begin{array} {cc}
0   & 1    \\
1   & 0
\end{array} \!\!
\right),
\label{eq7}
\end{equation}%
i.e. in this state there is a total transmission of the power from port 1 to port 2. The eigenvalues of this matrix are $s_{e}=1$ and $s_{o}=-1$. The corresponding normalized orthogonal eigenvectors ${\bf V}_{e}$ and ${\bf V}_{0}$ can be written as follows:
\begin{equation}
{\bf V}_{e} =\frac{1}{\sqrt{2}}\left(\!
\begin{array}{c}
1 \\
1
\end{array}\!
\right), \qquad {\bf V}_{o} = \frac{1}{\sqrt{2}}\left(\!
\begin{array}{c}
1 \\
\!\!\!-1
\end{array}\!
\right).
\label{eq8}
\end{equation}%
 The eigenvector ${\bf V}_{e}$ belongs to IRREP $A$ of the group $C_{s}$ describing the even excitation, and the vector ${\bf V}_{o}$ to IRREP $B$ corresponding to the odd excitation. The sum of ${\bf V}_{e}$ and ${\bf V}_{o}$ gives excitation of only port 1. The eigenvector-eigenvalue analysis which we shall use in the following, allows one to separate some constituents in the complex process of  the structure excitation and to simplify the identification of working modes of the devices. 
\section{Numerical Experiments with Eigenvectors}
\subsection{General Characteristics of Filter}
\begin{figure}[t]
\centering
\includegraphics[width=35pc]{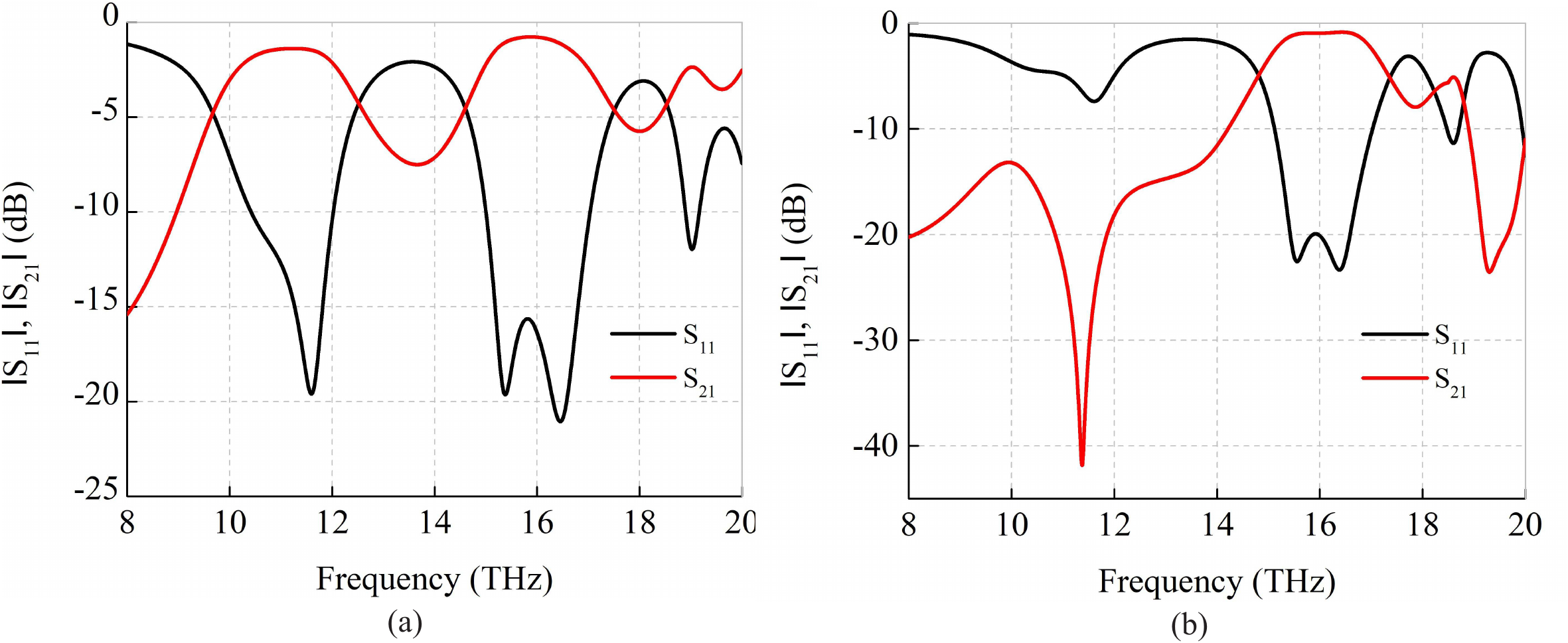}
\caption{(a) Frequency responses of collinear band-pass filter, b) of  filter with $90^\circ$ bending, $L$ = 900 nm, $R$ = 400 nm, $g$ = 2 nm, $w$ = 200 nm, $\epsilon_F$ = 0.5 eV and $\tau$ = 0.5 ps.}
\label{fig_RE90}
\end{figure}

 The working principle of operation of filter with the 90$^{\circ}$ bending is based on SPP waves in the graphene strips, which excite the dipole, quadrupole or hexapole resonance in the resonator.  
\begin{figure}[t]
\centering
\includegraphics[width=35pc]{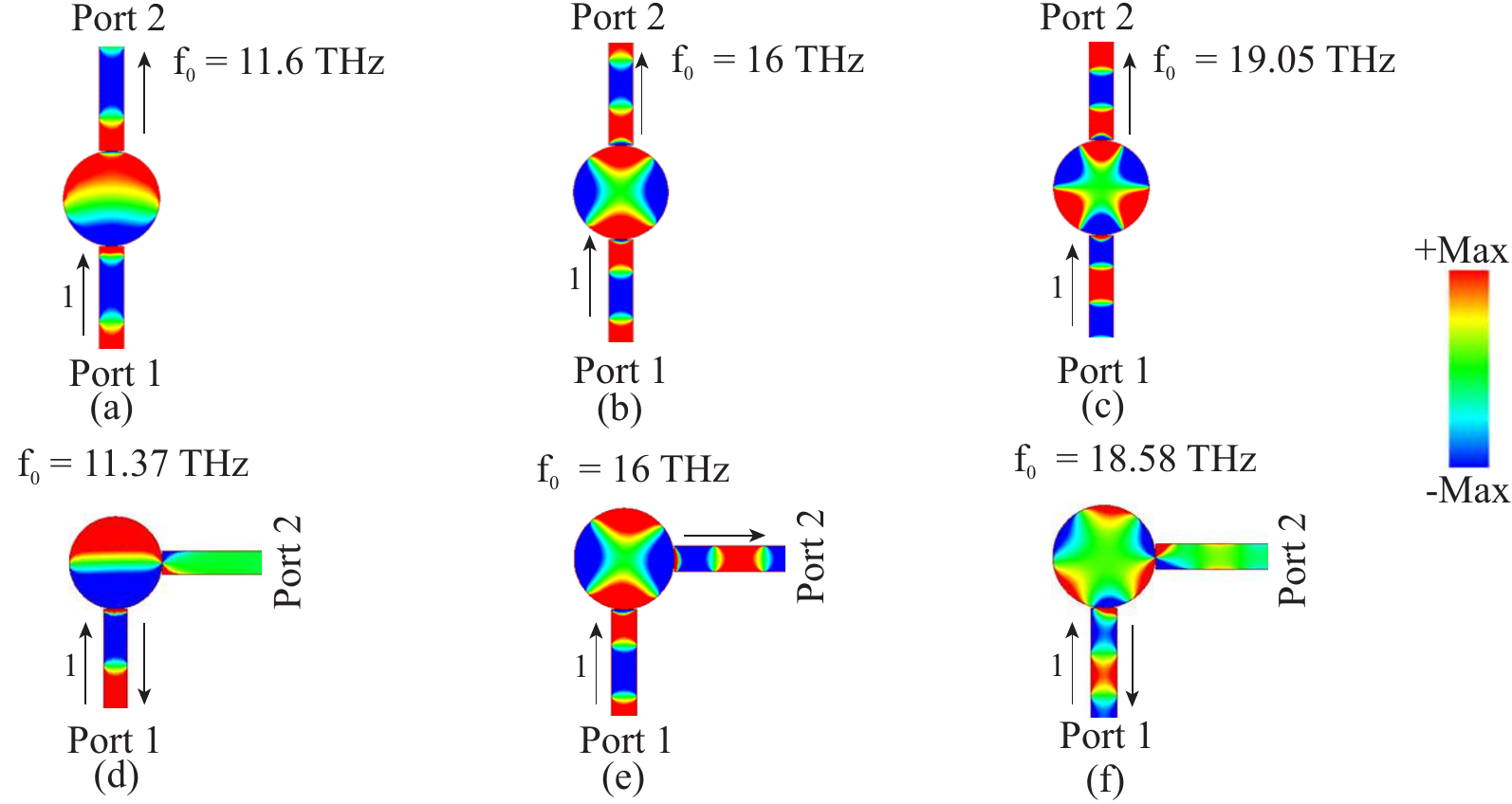}
\caption{$|E_z|$ field distribution in collinear band-pass filter and filter with 90$^{\circ}$ bending;  a) and d) dipole mode, b) and e) quadrupole mode, c) and f) hexapole mode, $L$ = 900 nm, $R$ = 400 nm, $g$ = 2 nm, $w$ = 200 nm,  $\epsilon_F$ = 0.5 eV and $\tau$ = 0.5 ps.}
\label{fig_Ez}
\end{figure}
 The structures of $|E_z|$ field in the discussed components are shown in Fig.\ref{fig_Ez} a, b, c for the collinear filter, and in 
Fig.\ref{fig_Ez} d, e, f for the filter with bending. The frequency characteristics of them are presented in Fig.\ref{fig_RE90}a and 
Fig.\ref{fig_RE90}b. In the collinear filter, the  three resonant modes define the pass-band regimes. In the filter with bending, there are three transmission extrema at the frequencies 11.37 THz (dipole mode, stop-band filter, the level of transmission is -43.6 dB), 16 THz (quadrupole mode, pass-band filter, the level of transmission is -0.73 dB) and 18.58 THz (hexapole mode, stop-band filter, the level of transmission is -10.2 dB). One should notice a very wide band and very high level of isolation of port 2 in the case of the dipole resonance. 
Notice also that the resonant frequencies of the corresponding modes in the upper and lower lines of Fig.\ref{fig_Ez} are very close.
\subsection{Stop-band Filter with Dipole Mode}
\begin{figure}[t]
\centering
\includegraphics[width=30pc]{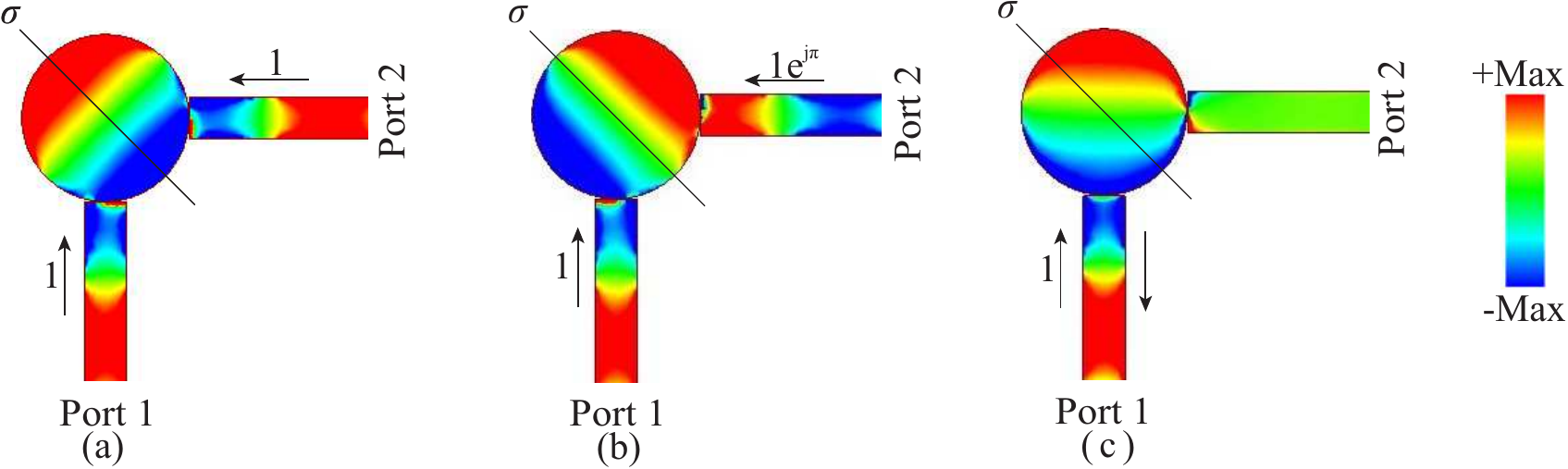}
\caption{Dipole resonance excitation: (a) even excitation, (b) odd excitation, (c) excitation of port 1, distribution of $|E_z|$ component at frequency 11.37 THz, $L$ = 900 nm, $R$ = 400 nm, $g$ = 2 nm, $w$ = 200 nm, $\epsilon_F$ = 0.5 eV and $\tau$ = 0.5 ps.}
\label{fig_Fig2b}
\end{figure}

 Firstly, we apply to the dipole mode of the disk resonator. The ideal dipole mode in the uncoupled resonator is described by the symmetry $C_{2v}$. However, connection of two nanostrips to the resonator reduces the symmetry of the structure to $C_{s}$ and this leads to a certain distortion of the dipole. We shall consider two possible excitations: by vector ${\bf V}_{e}$ shown in Fig.\ref{fig_Fig2b} a, and by vector ${\bf V}_{o}$ demonstrated in Fig.\ref{fig_Fig2b} b. 
  One can see  orthogonal orientation of the dipole standing waves in these two cases. The local fields near the waveguide-resonator  junctions in Fig.\ref{fig_Fig2b} a and Fig.\ref{fig_Fig2b} b correspond to the schemes of Fig.\ref{fig_Appendix2} c
  and Fig.\ref{fig_Appendix2} d, respectively.

 In our component, the input and output ports are oriented at $90^\circ$. In case of the dipole excitation of port 1, the absence of transmission to port 2 is stipulated by the orthogonal structure of the resonator field and the field of the waveguide eigenmode. As it was mentioned already, in the input and output graphene waveguides we consider the principal SPP wave which is symmetric with respect to the plane of symmetry of the graphene strip $\sigma_{wi}$, see Fig.\ref{fig_Fig2b} a, b, c. The nodal plane of the resonator standing wave is in the plane of symmetry of the output waveguide and this leads to a ``natural" isolation of the output port. 
\subsection{Pass-band Filter with Quadrupole Mode}
 Different types of excitation of the quadrupole mode are demonstrated schematically in 
Fig.\ref{fig_Fig3b}.
\subsubsection{Odd Excitation of Quadrupole Mode}
\begin{figure}[t]
\centering
\includegraphics[width=30pc]{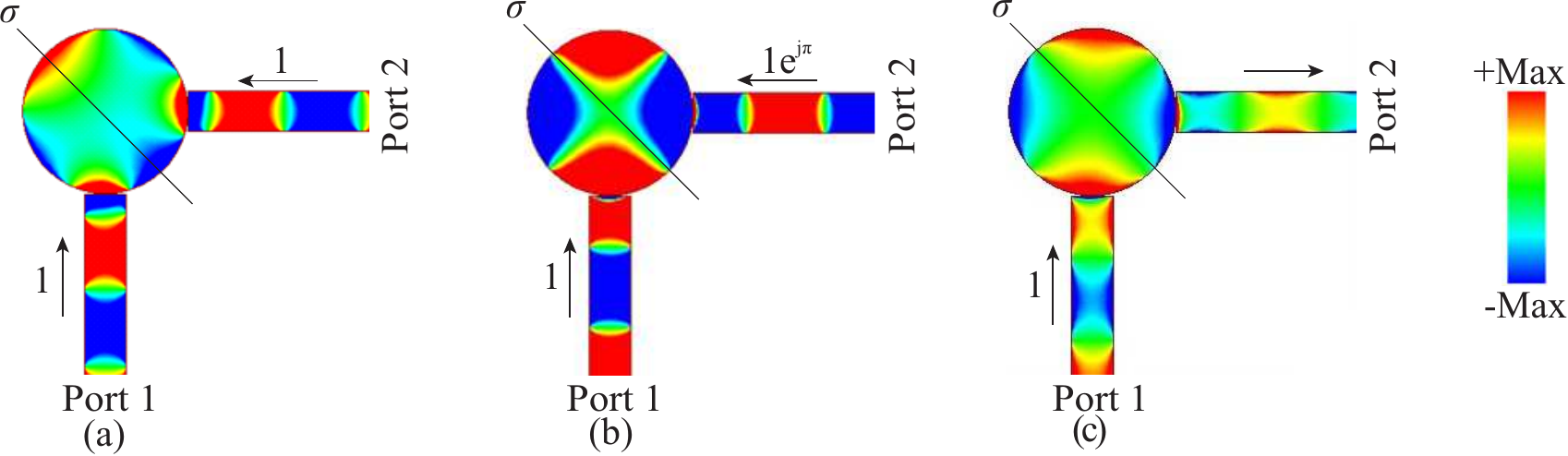}
\caption{Excitation around quadrupole resonance: a) even excitation, b) odd excitation, Q1 mode,  c) excitation of port 1 with Q1 mode at frequency 16.42 THz;  $R$ = 400 nm, $g$ = 2 nm, $w$ = 200 nm, $L$ = 900 nm, $\epsilon_F$ = 0.5 eV and $\tau$ = 0.5 ps. }
\label{fig_Fig3b}
\end{figure}

 We show in Fig.\ref{fig_Fig3b} b the case of odd excitation of the two-port with the quadrupole resonance which we shall call further quadrupole 1 (Q1). The $|E_z|$ field of Q1 in the resonator is anti-symmetric with respect to the plane 
$\sigma$. 
\subsubsection{Even Excitation of Quadrupole Mode}
For the even excitation by eigenvectors, the quadruple mode which will be called further Q2, can not be excited in the resonator because the fields in the waveguides and in the resonator are orthogonal. In this case one can say, that the quadrupole mode Q2 belongs to the so-called dark or trapped modes \cite{Dmitriev}. Notice, that in our numerical experiment of the even excitation, one can see in the resonator a symmetric field (in Fig.\ref{fig_Fig3b} a) but corresponding to another mode, namely, to the hexapole one. This hexapole mode is highly distorted because of the low symmetry of the structure. 
\subsubsection{Excitation of Port 1}
 The frequency response of the filter with quadrupole mode for excitation of port 1 has two dims (see Fig.\ref{fig_RE90} b). Numerical analysis shows that these dims corresponds to the same Q1 mode. The presence of these  dims can be explained as follows. In case of excitation of  port 1, the symmetry of the problem $C_s$ is broken by non-symmetrical excitation because only port 1 is excited. As a result, the mode Q2 can be excited in the resonator, i.e. this mode should  be called now a quasi-dark one. 
\begin{figure}[t]
\centering
\includegraphics[width=28pc]{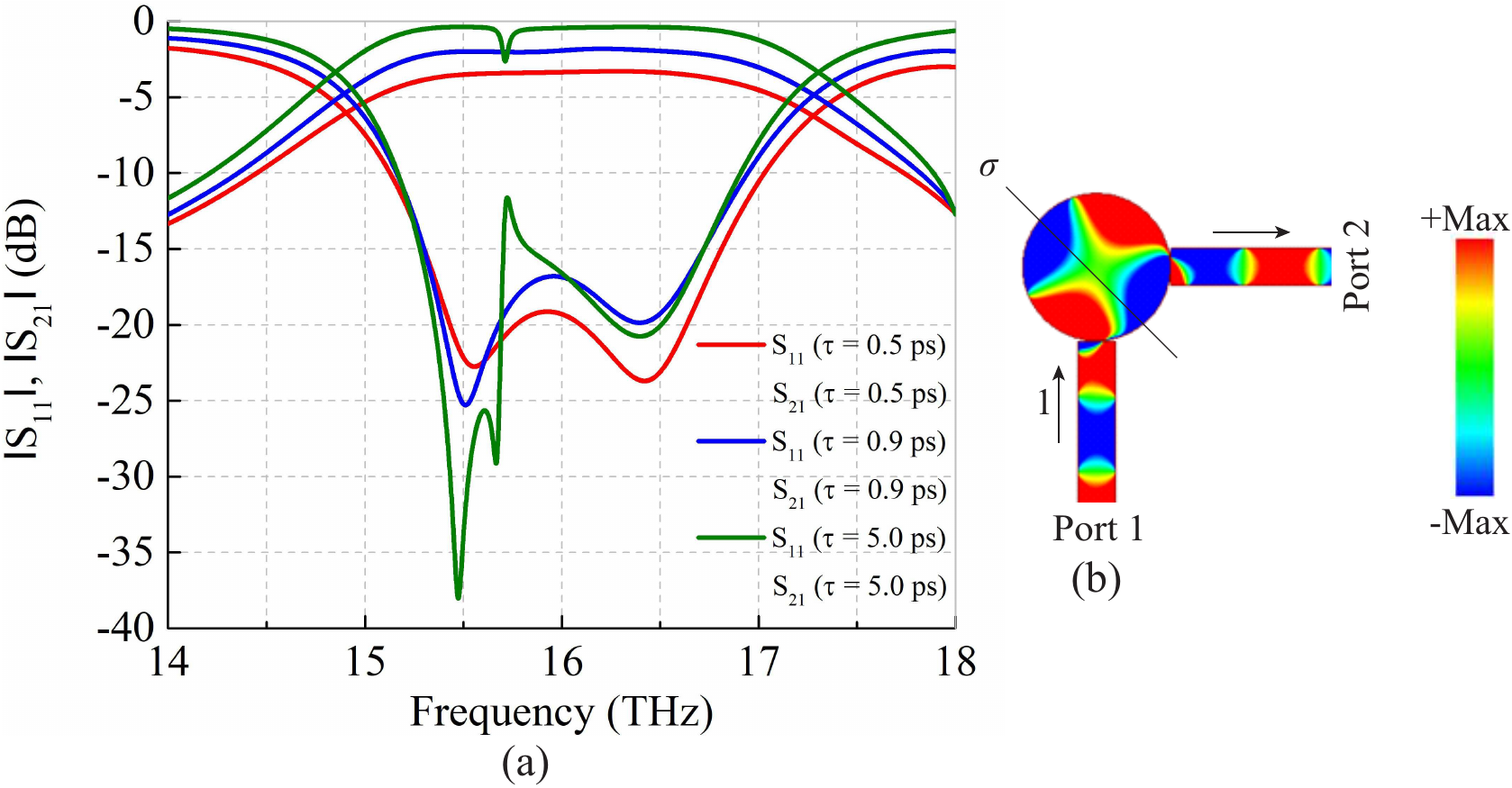}
\caption{(a) Frequency response for quadrupole mode for different $\tau$ values demonstrating presence of quasi-dark mode Q2; (b) Distribution of $|E_z|$ component of mode Q2, $\tau$ = 5 ps, for frequency 15.74 THz, $R$ = 400 nm, $g$ = 2 nm, $L$ = 900 nm, $w$ = 200 nm, $\epsilon_F$ = 0.5 eV.}
\label{fig_TAU5}
\end{figure}

 In a more wide context, the mode Q2 can be called also a symmetry-protected bound state in the continuum (BIC) \cite{Koshelev, Shaimaa, JW} which is characterized by a very high quality factor and specific frequency responses presenting a sharp peak-and-through curve as shown in Fig.\ref{fig_TAU5} a. It should be stressed a peculiarity  of our case: the resonance Q1 with a low quality factor (continuum) and the resonance Q2 with the high quality factor (i.e. quasi-dark mode) are the same eigenmodes of the resonator, namely, the quadrupole ones. The difference between them is orientation of the field structure, i.e. a certain rotation (ideally, by $45^\circ$) around the z-axis of one field structure  with respect to the other (compare Fig.\ref{fig_Fig3b} b and Fig.\ref{fig_TAU5} b).
 
 Comparing Fig.\ref{fig_TAU5} b and Fig.\ref{fig_Appendix2} c of Appendix, one can see that the structure of $|E_z|$ field around the gap $g$ corresponds to the symmetry $C_s \otimes C_s$.
\subsection{Stop-band Filter with Hexapole Mode}
\begin{figure}[t]
\centering
\includegraphics[width=30pc]{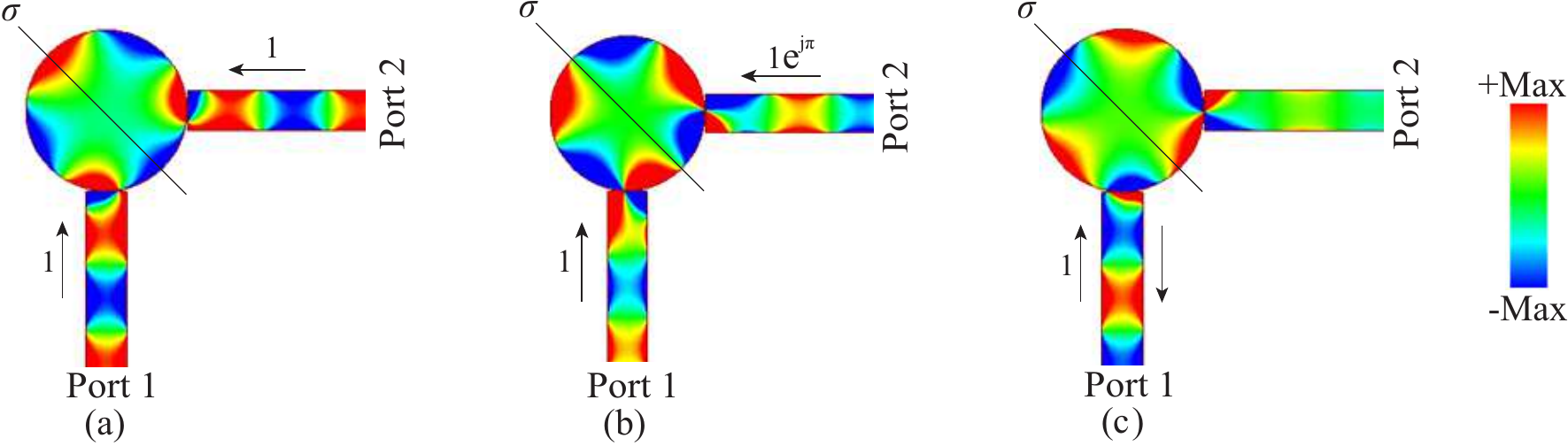} 
\caption{Hexapole resonance excitation: a) even excitation, b) odd excitation, c) excitation of port 1; $R$ = 400 nm, $g$ = 2 nm, $w$ = 200 nm,  $L$ = 900 nm, $\tau$ = 0.5 ps, $\epsilon_F$ = 0.5 eV and frequency 15.58 THz.}
\label{fig_Fig4b}
\end{figure}

The results of numerical calculus for the  even excitation, odd excitation and excitation of port 1 for hexapole mode are presented in Fig.\ref{fig_Fig4b} a, b, c.
The field structure of the hexagons is slightly distorted. As in case of dipole resonance (Fig.\ref{fig_Fig2b} c), one can see in Fig.\ref{fig_Fig4b} c that the nodal plane of the resonator standing wave is in the plane of symmetry of the output waveguide and this leads again to a ``natural" isolation of the output port.
\section{Simulation Results for Switch}
For the switch projects, one can consider two variants of the resonant modes combination: dipole and quadrupole (D+Q) or quadrupole and hexapole (Q+H) ones. In particular, in D+Q variant combining the stop-band filter properties of the dipole resonance and pass-band filter properties of the quadrupole resonance, one can realize a switch.  A dislocation of the quadrupole resonance frequency to the dipole one can be achieved by changing the graphene Fermi energy. One can consider also dislocation of the dipole resonance frequency to the quadrupole one.  Analogously, a combination of quadrupole and  hexapole resonances (Q+H variant) can also provide switching mechanism. 

From (\ref{eq3}) one  obtains:
\begin{equation}
\epsilon_F =\omega_{c}^2 R (1+\varepsilon_1)/(na_nA).
\label{eq9}
\end{equation}%
This formula shows that for a given $\omega_{c}$, $\varepsilon_1$ and the chosen $R$, the product $na_n\epsilon_F$ is constant.  This allows one to relate the necessary values of  $\epsilon_F$ in the switch projects. If we take $a_n\approx 1$, for the D+Q switch $\epsilon_{FD}$ is approximately two times higher than $\epsilon_{FQ}$  for the quadrupole mode,  i.e. $\epsilon_{FD}=2\epsilon_{FQ}$.  If we consider a maximum possible value of $\epsilon_F=1$, the maximum admissible 
value for $\epsilon_{FD}$ will be 0.5. For the Q+H switch,  the relation between $\epsilon_{FQ}$ of the quadrupole mode and  $\epsilon_{FH}$ of the hexapole mode is $\epsilon_{FQ}=1.5\epsilon_{FH}$. 
\subsection{Example of D-Q Switch}
\begin{figure}[t]
\centering
\includegraphics[width=28pc]{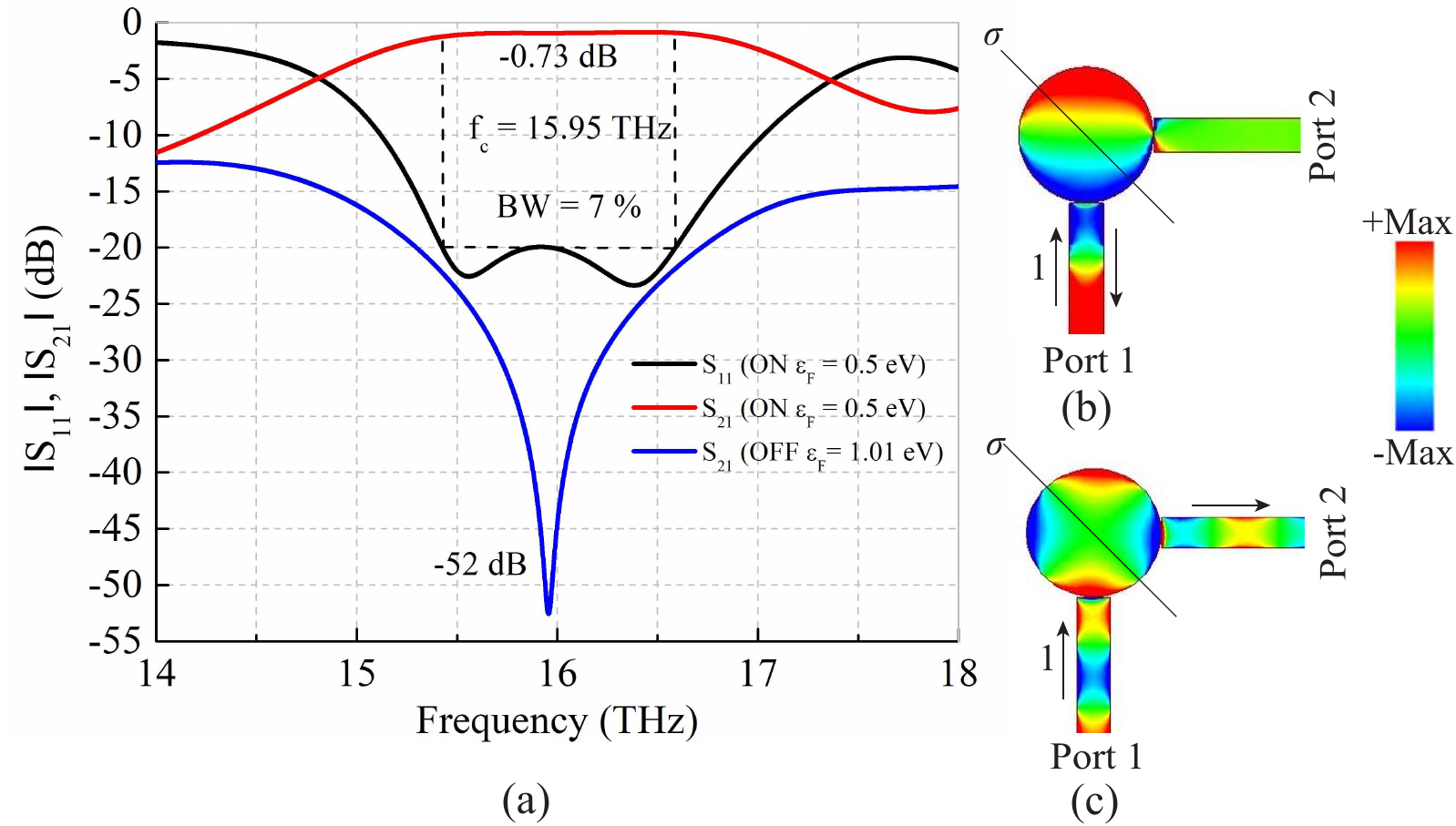}
\caption{(a) Frequency responses of D+Q  switch in ON ($\epsilon_F $ = 0.5 eV) and OFF ($\epsilon_F$ = 1.01 eV) states; (b) and (c) $|E_z|$ field distribution for frequency 15.95 THz,  $R$ = 400 nm, $L$ = 900 nm, $g$ = 2 nm, $w$ = 200 nm, $\tau$ = 0.5 ps.}
\label{fig_ONOFF4}
\end{figure}

 Distribution of $|E_z|$ field in switch with D+Q mechanism is shown in Fig.\ref{fig_ONOFF4} b and Fig.\ref{fig_ONOFF4} c, with dipole mode corresponding to the state OFF, and quadrupole mode for the state ON. The frequency characteristics of the switch are given in Fig.\ref{fig_ONOFF4} a. 
 The Fermi energy of graphene in this case is switched from 0.5 eV to 1.01 eV. In Fig.\ref{fig_ONOFF4} a one can see that the device operates at the center frequency 15.95 THz. For the Fermi energy $\epsilon_{FQ}$ = 0.5 eV, we have the ON state with insertion losses of -0.73 dB and reflection of -23 dB for the quadrupole mode, while for the dipole mode in OFF state, the  Fermi energy $\epsilon_{FD}$ = 1.01 eV, the isolation is about -52 dB and the bandwidth BW = 7$\%$, considering the reflection level of -20 dB. Thus the working bandwidth of the switch is about 1 THz with the central frequency $f_0$ = 15.95 THz. 
 The switch modulation depth calculated by considering an isolation level of -20 dB and transmission of -1 dB is $MD = (T_\textit{ON} - T_\textit{OFF})/T_\textit{ON}$,
where $T_\textit{ON}$ = 0.79 and $T_\textit{OFF}$ = 0.01 are the magnitudes of the transmittances in the ON and OFF states, respectively. Thus, we obtain MD = 0.99. 

The choice of the D+Q or Q+H combination in switch design  depends on the frequency band of the switch and on desired parameters of it in the states ON and OFF.  Finally, it should be stressed that it is impossible to realize the suggested mechanism of switching in the known collinear structures because all their resonances correspond to the pass-band regime.
\section{Discussion}
 The shift of the Fermi energy and, as a result, a change of the graphene surface conductivity by
application of a DC voltage is a principal mechanism of switching. The DC voltage can be applied between the graphene sheet and a very thin polysilicon layer with relatively high conductivity which is used as a gate electrode \cite{Gomez}, \cite{Yan}. A choice of the dielectric material between graphene and the gate electrode (such as for example, HfO$_2$, TiO$_2$, Al$_2$O$_3$ and ion gel gate dielectrics) which allow one to change the Fermi energy up to 1.3 eV  without  voltage breakdown is  discussed in \cite{Fus}. We have shown above that the resonance frequency depends mostly in the radius of the graphene resonator and on Fermi energy of graphene. The width of the graphene strips $w$ and the gap $g$ between the resonator and the strips shown in Fig.\ref{fig_Fig12} also affect the resonance frequencies but to a lesser degree. This problem has been discussed in detail in \cite{circulator} for the case of graphene three-port circulators based on disk resonator. 

Thus, we have used  here dependence of the graphene disc resonance  on the Fermi energy in two different manners. Firstly, to produce switch, and secondly, to adjust dynamically  the working frequency of the discussed filters and switches.

 In this article, we have chosen the graphene relaxation time $\tau$ = 0.5 ps. Today this parameter can be used in practice to develop guided wave devices and it is possible to get realistic results due to the quality of graphene we currently have. Under some special conditions by using, for example, new materials such as hexagonal boron nitride \cite{BoronNitrid} as substrate, a higher relaxation time for graphene can be achieved and this reduces the insertion losses of the devices. With very rapid progress in graphene technology, it is hoped that in the coming years the quality of graphene will be improved further. 
 
 A theoretical estimation of the switching energy of the graphene switches in mid-infrared band gives a very low value of about 40.0 fJ/bit \cite{SwtchEnergy2}. Experimentally obtained switching speed is 6 kHz, see \cite{Gomez}, \cite{Speed1}.
\section{Conclusions}
In this work we have suggested, analysed and confirmed by numerical simulations a possibility of realization of  novel plasmonic graphene-based tunable  filters and switches.  We have discussed  a new mechanism of switching which is based on combination of the stop-band  and pass-band  filter properties of the graphene resonator. Switching the graphene Fermi energy by the electrostatic field effect, the central frequency of the pass-band  can be dislocated until coincidence with that of the stop-band. This corresponds to transition from the OFF state to the ON state. Thus, a combination in the transmission response of the filter of a high peak for one value of Fermi energy (pass-band, high transmission, regime ON) and a deep valley (stop-band, low transmission, regime OFF) for another value of Fermi energy  allows one to create a switch with high modulation depth using combinations of quadrupole-dipole or quadrupole-hexapole resonances in the graphene disc. In the analysis of the devices we have used the method of even and odd excitation which allows one to separate the effects of different modes on the performance of the switches and to facilitate their analysis.

 The center frequency of the device is defined by the radius of the disc resonator.  Varying DC voltage between the graphene sheet and a gate electrode  provides change in the  Fermi energy and consequently, a frequency shift of the filter characteristics, i.e. dynamic control of filter and switch. The radius $R$ of the resonator of the proposed devices is  $ 0.53\lambda_{SPP}\approx 0.02\lambda_0$. Thus, the footprints of the suggested components are  very small and this is explained firstly, by small wavelength of the SPP and secondly, by resonant mechanism of the devices. From the point of view of time and energy consumption, one can expect that  these parameters in the suggested switches will be similar to those of the known switches based on the electro-optical effect. The results of this study can be used to design filters, switches and modulators operating in THz and FIR regions.

\section*{Acknowledgements}
This work was supported by the Brazilian Agency National Counsel of Technological and Scientific Development CNPq, Federal University of Para (UFPA) and Federal Rural University of Amazonia (UFRA).

\section*{Appendix}
\begin{table}[!t]
\caption{Irreducible representations of the group $C_s$.}
\centering
\begin{tabular}{|c|c|c|}\hline
$C_s$ & $e$ & $\sigma$ \\\hline
$A$ & 1 & 1 \\\hline
$B$ & 1 & -1 \\\hline
\end{tabular}
\end{table}
\begin{figure}[t]
\centering
\includegraphics[width=21pc]{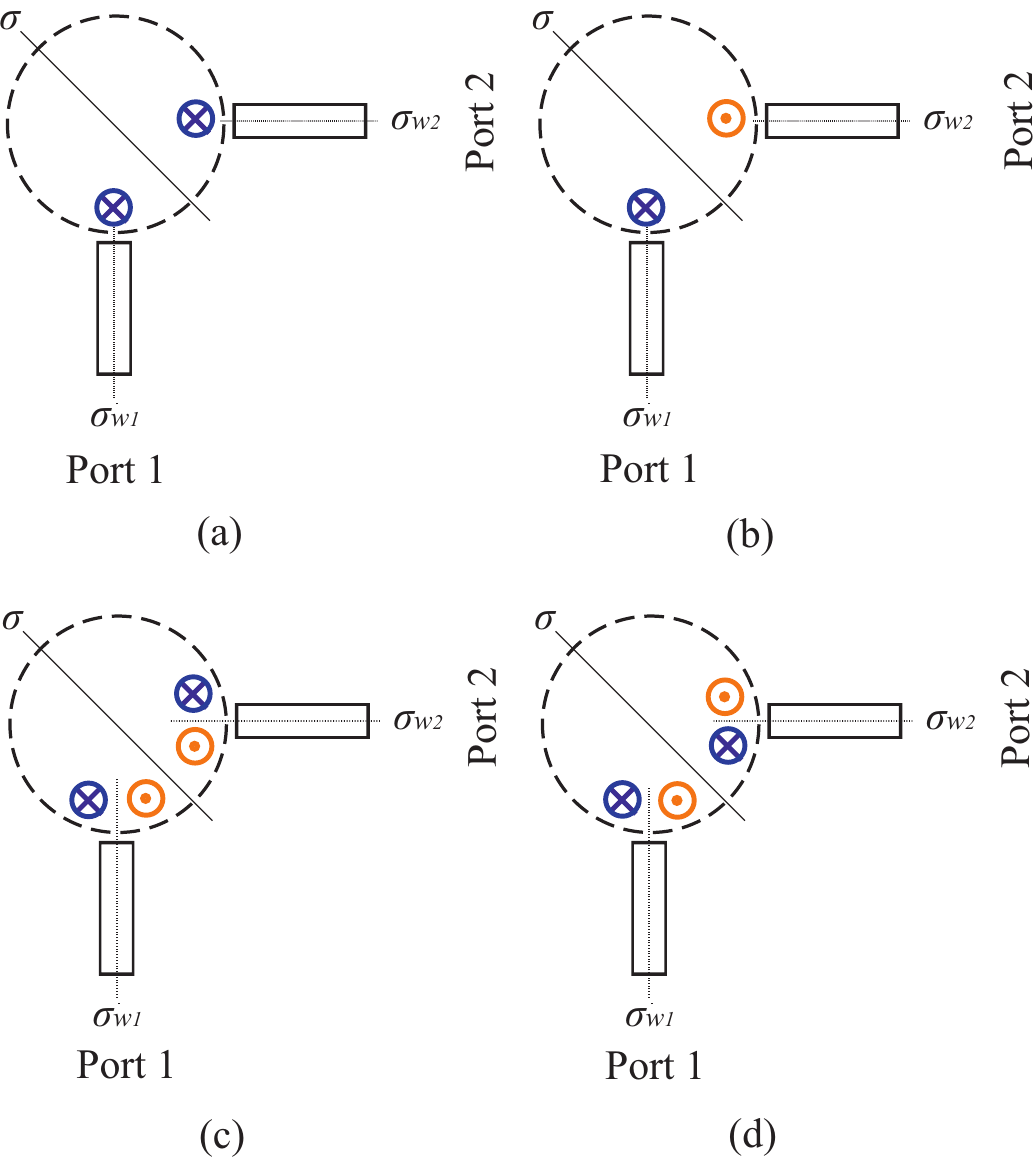}
\caption{Schematics of a) even and b) odd field structure in symmetry $C_s$; c) even and d) odd structure of field in case of symmetry $C_s \otimes C_s$.}
\label{fig_Appendix2}
\end{figure}
%


\end{document}